\documentclass[pra,aps,superscriptaddress,floatfix,twocolumn]{revtex4-1}
\usepackage{amsmath,amssymb,multirow,epsfig,bm,pifont}
\usepackage[linkcolor=black,citecolor=black,urlcolor=blue,colorlinks=true,linktocpage=true]{hyperref}
\usepackage{graphicx}
\usepackage{epsfig}
\usepackage{bm}
\usepackage{physics}
\usepackage{tabularx}
\usepackage{makecell}

\newcommand{\beq}{\begin{equation}}
\newcommand{\eeq}{\end{equation}}
\newcommand{\bea}{\begin{eqnarray}}
\newcommand{\eea}{\end{eqnarray}}

\newcommand{\la}{\langle}
\newcommand{\ra}{\rangle}

\begin{document}

\title{Third- and fourth-order virial coefficients of harmonically trapped fermions in a semiclassical approximation}

\author{K. J. Morrell}
\affiliation{Department of Physics and Astronomy, University of North Carolina, Chapel Hill, NC, 27599, USA}
\author{C. E. Berger}
\affiliation{Department of Physics and Astronomy, University of North Carolina, Chapel Hill, NC, 27599, USA}
\author{J. E. Drut}
\affiliation{Department of Physics and Astronomy, University of North Carolina, Chapel Hill, NC, 27599, USA}

\begin{abstract}
Using a leading-order semiclassical approximation, we calculate the third- and fourth-order virial coefficients
of nonrelativistic spin-1/2 fermions in a harmonic trapping potential in arbitrary spatial dimensions,
and as functions of temperature, trapping frequency and coupling strength.
Our simple, analytic results for the interaction-induced changes $\Delta b_3$ and $\Delta b_4$
agree qualitatively, and in some regimes quantitatively, with previous numerical calculations for 
the unitary limit of three-dimensional Fermi gases.
\end{abstract}

\date{\today}

\maketitle

\section{Introduction}

The properties of fermions at finite temperature and density are in part governed by the dimensionless product $\beta \mu$, where 
$\beta = 1/(k_B T)$ is the inverse temperature and $\mu$ is the chemical potential. Typically, the region $\beta \mu \simeq 0$ 
displays a crossover between quantum and classical physics, while $z = e^{\beta \mu} \ll 1$ indicates a dilute limit where 
the thermodynamics is given by the virial expansion, which expands a given physical quantity in powers of $z$. 
Since $\mu$ is coupled to the particle number $N$, the virial expansion at order $N$ contains the physics of the $N$-body problem.
In the simplest case, the coefficients $b_n$ of the virial expansion determine the pressure, density, and compressibility, as well as
other elementary thermodynamic quantities such as energy and entropy. The change in $b_n$ due to interactions is usually denoted
$\Delta b_n$.

The previous work of Ref.~\cite{ShillDrut} calculated the third- and fourth-order virial coefficients $\Delta b_3$ and $\Delta b_4$, respectively, 
at leading-order (LO) in a semiclassical lattice approximation (SCLA), of homogeneous spin-$1/2$ fermions in arbitrary dimension. The follow-up work of Ref.~\cite{HouEtAl} extended those results up to $\Delta b_7$, while Ref.~\cite{HouDrut} carried out calculations up to next-to-next-to-leading order 
in the SCLA for up to $\Delta b_5$. 
In this brief work we provide another piece of the puzzle by generalizing the calculations of Ref.~\cite{ShillDrut} to 
systems in a harmonic trap of frequency $\omega$. We present our derivations with intermediate steps in 
detail and give analytic formulas for $\Delta b_3$ and $\Delta b_4$ as functions of $\beta \omega$ in arbitrary spatial dimension $d$.
Our results, which will be given in terms of $\Delta b_2$, are thus also functions of the coupling strength.

\section{Hamiltonian and formalism}

As our focus is on systems with short-range interactions, such as dilute atomic gases or dilute neutron matter, 
the Hamiltonian reads
\beq
\hat H = \hat H_0 +  \hat V_\text{int},
\eeq
where
\beq
\hat H_0 = \hat T + \hat V_\text{ext},
\eeq
and
\bea
\label{Eq:T}
\hat T \!=\! \sum_{s = 1,2} {\int{d^d x\,\hat{\psi}^{\dagger}_{s}({\bf x})\left(-\frac{\hbar^2\nabla^2}{2m}\right)\hat{\psi}_{s}({\bf x})}},
\eea
is the kinetic energy,
\bea
\label{Eq:Vext}
\hat V_\text{ext} \!=\! \frac{1}{2} m \omega^2 \int{d^d x}\, {\bf x}^2 \,( \hat{n}_{1}({\bf x}) + \hat{n}_{2}({\bf x})) ,
\eea
is the spherically symmetric external trapping potential, and
\bea
\label{Eq:Vint}
\hat V_\text{int} \!=\! - g_{dD}\! \int{d^d x\,\hat{n}_{1}({\bf x})\hat{n}_{2}({\bf x})},
\eea
is the interaction.

In the above equations, the field operators $\hat{\psi}_{s}, \hat{\psi}^{\dagger}_{s}$ correspond to particles of species $s=1,2$, 
and $\hat{n}_{s}({\bf x})$ are the coordinate-space densities. For the remainder of this work, we will set $\hbar = k_\text{B} = m = 1$.

\subsection{Thermodynamics and the virial expansion}

The equilibrium thermodynamics of our quantum many-body system can be captured by the grand-canonical partition function,
namely
\beq
\mathcal Z = \tr \left[ e^{-\beta (\hat H -\mu \hat N)}\right] = e^{-\beta \Omega}
\eeq
where $\beta$ is the inverse temperature, $\Omega$ is the grand thermodynamic potential, $\hat N$ is the total particle number operator, 
and $\mu$ is the overall chemical potential (we will not consider polarized systems in this work).

As the calculation of $\mathcal Z$ is a formidable problem in the presence of interactions, we resort to approximations or numerical
evaluations in order to access the thermodynamics. To that end, in this work we will use the virial expansion, which is an expansion around the dilute limit
$z\to 0$, where $z=e^{\beta \mu}$ is the fugacity, i.e. it is a low-fugacity expansion (see Ref.~\cite{VirialReview} for a review on recent applications
of the virial expansion to ultracold atoms). The coefficients accompanying the powers of $z$ in the expansion $\Omega$ are the virial coefficients $b_n$:
\beq
-\beta \Omega = \ln {\mathcal Z} = Q_1 \sum_{n=1}^{\infty} b_n z^n,
\eeq
where $Q_1$ is the one-body partition function. Using the fact that $\mathcal Z$ is itself a 
sum over canonical partition functions $Q_N$ of all possible particle numbers $N$, namely
\beq
\mathcal Z = \sum_{N=0}^{\infty} z^N Q_N,
\eeq 
we obtain expressions for the virial coefficients
\bea
b_1&=& 1,\\
b_2 &=& \frac{Q_2}{Q_1} - \frac{Q_1}{2!},\\
b_3 &=& \frac{Q_3}{Q_1} - b_2 Q_1  - \frac{Q_1^2}{3!},\\
b_4 &=& \frac{Q_4}{Q_1}  -  \left(b_3 + \frac{b_2^2}{2}\right) Q_1 -b_2\frac{Q_1^2}{2!}  - \frac{Q_1^3}{4!},
\eea
and so on. In this work we will not pursue the virial expansion beyond $b_4$. 
The $Q_N$ can themselves be written in terms of the partition functions $Q_{a,b}$ for $a$ particles of type 1 and $b$ particles of type 2:
\bea
Q_1 &=& 2 Q_{1,0},\\
Q_2 &=&  2 Q_{2,0} + Q_{1,1},\\
Q_3 &=&  2 Q_{3,0} + 2 Q_{2,1},\\
Q_4 &=& 2 Q_{4,0} +  2 Q_{3,1} + Q_{2,2},
\eea
and so on for higher orders. In the absence of intra-species interactions, only $Q_{1,1}$, $Q_{2,1}$, $Q_{3,1}$, and $Q_{2,2}$  
are affected, such that the change in $b_2$, $b_3$, and $b_4$ due to interactions is entirely given by
\bea
\!\Delta b_2\! &=&\! \frac{\Delta Q_{1,1}}{Q_1}, \\
\!\Delta b_3\! &=&\! \frac{2 \Delta Q_{2,1}}{Q_1} \!-\!\Delta b_2 Q_1, \\
\!\Delta b_4\! &=&\! \frac{2 \Delta Q_{3,1} \!+\! \Delta Q_{2,2}}{Q_1}\!-\! \Delta\!\left(b_3 + \frac{b_2^2}{2}\right)\! Q_1\!-\!\frac{\Delta b_2}{2} Q_1^2.
\eea
To calculate $\Delta Q_{m,n}$, we implement a semiclassical approximation, as described in the next section. Once we obtain the virial 
coefficients, one may rebuild the grand-canonical potential $\Omega$ to access the thermodynamics of the system as a function of
the various parameters. 

In order to connect to the {\it physical} parameters of the systems at hand, we will use the value of $\Delta b_2$
as a renormalization condition by relying on the exact answers as functions of $\beta \omega$ and the physical coupling $\lambda$.
These exact answers are not always known analytically, but they can easily be obtained numerically by solving the two-body problem
of interest.

Although in this work we will focus on systems in a harmonic trap, thus far the identities presented in this section are more general.
As a reference for the trapped case, we present here the calculation of the noninteracting virial coefficients for arbitrary $\beta \omega$.
[We note that such a calculation, while simple, does not appear in the literature.]
Starting from the logarithm of the noninteracting partition function in $d$ spatial dimensions, we have, for two fermion species,
\beq
\ln \mathcal Z = 2 \sum_{\bf n} \ln \left(1 + z e^{- \beta d/2} \prod_{i=1}^{d} e^{-\beta \omega n_i} \right).
\eeq
Expanding in powers of $z$ on both sides, and switching the order of the sums, we obtain
\bea
Q_1 \sum_{k=1}^{\infty} b^{0}_k z^k &=& 2\sum_{k = 1}^{\infty} z^k \frac{(-1)^{k+1}}{k} e^{- \beta \omega d k/2} \left( \sum_{n=0}^{\infty} e^{-\beta \omega k n} \right)^d 
\nonumber \\
&=& 2\sum_{k=1}^{\infty} z^k \frac{(-1)^{k+1}}{2^d k} \left( \frac{1}{\sinh (\beta \omega k /2)}\right)^d.
\eea
To identify the noninteracting virial coefficients $b^{0}_n$, we need $Q_1$:
\bea
\label{Eq:Q1}
Q_1&=&2\sum_{\bf n}e^{-\beta E_n}=2e^{-\beta \omega d /2}\bigg(\frac{1}{1-e^{-\beta \omega}}\bigg)^d\\ 
&=& 2\left( \frac{1}{2 \sinh(\beta \omega /2)} \right)^{d}.
\eea
Thus, the virial coefficients of a trapped noninteracting spin-1/2 Fermi gas in $d$ dimensions are
\beq
\label{Eq:NonIntbn}
b_n^{0} = \frac{(-1)^{n+1}}{n} \left( \frac{\sinh(\beta \omega /2)}{\sinh (\beta \omega n /2)}\right)^d.
\eeq
Notably, in the limit $\beta \omega \ll 1$, we obtain
\beq
b_n^{0} \to (-1)^{n+1} \left( \frac{1}{n}\right)^{d+1},
\eeq
which agrees in $d=3$ with the local density approximation result quoted in Ref.~\cite{VirialReview}.
The simple result of Eq.~(\ref{Eq:NonIntbn}) should be a textbook calculation, but it does not appear elsewhere,
to the best of our knowledge.
Note that for the homogeneous (i.e. untrapped) system, the noninteracting virial coefficients in $d$ dimensions are
\beq
b_n^{0, \text{hom}} = (-1)^{n+1} \left( \frac{1}{n}\right)^{\frac{d}{2}+1},
\eeq
such that $b_n^{0} = b_n^{0, \text{hom}} n^{-\frac{d}{2}}$ for $\beta \omega \ll 1$.

\subsection{Semiclassical lattice approximation}

To calculate the interaction-induced change $\Delta Q_{m,n}$, we implement an approximation 
which consists in keeping the leading term in the commutator expansion:
\beq
e^{-\beta (\hat H_0 + \hat V_\text{int})} = e^{-\beta \hat H_0}e^{-\beta \hat V_\text{int}}\times e^{-\frac{\beta^2}{2} [\hat H_0, \hat V_\text{int}]}\times \dots,
\eeq
where the higher orders involve exponentials of nested commutators of $\hat H_0$ with $\hat V_\text{int}$. Thus, the leading order 
in this expansion consists in setting $[\hat H_0 , \hat V_\text{int}] = 0$, which corresponds to a semiclassical approximation.
Another way to see this approximation is in terms of a Trotter-Suzuki factorization, i.e.
\beq
e^{-\beta (\hat H_0 + \hat V_\text{int})} = \lim_{n \to \infty} \left( e^{-\beta \hat H_0 / n}e^{-\beta \hat V_\text{int}/ n} \right)^n,
\eeq
where the $n=1$ case is the leading order we pursue in this work and higher orders can be defined by increasing $n$.

\subsection{Example: Calculation of $\Delta Q_{1,1}$ and $\Delta b_2$}

In the approximation proposed above, the two-particle problem is analyzed as follows:
\bea
Q_{1,1} &=& \text{Tr}\left[e^{-\beta\hat{H}_0}e^{-\beta\hat{V}_\text{int}}  \right]  \nonumber \\
&=& \sum_{{\bf k}_{1},{\bf k}_{2},{\bf x}_{1},{\bf x}_{2}}\la {\bf k}_{1} {\bf k}_{2} | e^{-\beta \hat{H}_0} | {\bf x}_{1}{\bf x}_{2} \ra \la {\bf x}_{1}{\bf x}_{2} | e^{-\beta\hat{V}_\text{int}} |{\bf k}_{1} {\bf k}_{2} \ra \nonumber \\
&=& \sum_{{\bf k}_{1},{\bf k}_{2},{\bf x}_{1},{\bf x}_{2}} e^{-\beta(E_{{\bf k}_{1}} + E_{{\bf k}_{2}})} M_{{\bf x}_{1},{\bf x}_{2}} |\la {\bf k}_{1} {\bf k}_{2} | {\bf x}_{1}{\bf x}_{2} \ra|^{2},
\eea
where we have inserted complete sets of states in coordinate space $\{| {\bf x}_1 {\bf x}_2 \rangle\}$ and in the basis $|{\bf k}_1 {\bf k}_2 \rangle $ of eigenstates of 
$\hat H_0$, whose single-particle eigenstates $| {\bf k} \rangle$ have eigenvalues $E_{\bf k}$.
We have also made use of the fact that $\hat V_\text{int}$ is diagonal in coordinate space, such that
\beq
M_{{\bf x}_{1},{\bf x}_{2}} = 1 + C \delta_{{\bf x}_1,{\bf x}_2},
\eeq
where $C = (e^{\beta g_{dD}} - 1) \ell^d$ and $\ell$ is an ultraviolet length scale to be defined by our renormalization condition 
(see below).

Thus,
\beq
\label{Eq:DQ11}
\Delta Q_{1,1} = C \sum_{{\bf k}_{1},{\bf k}_{2}} e^{-\beta(E_{{\bf k}_{1}} + E_{{\bf k}_{2}})} |\phi_{{\bf k}_1}({\bf x})|^2|\phi_{{\bf k}_2}({\bf x})|^2,
\eeq
and we will use normalized single-particle wavefunctions in cartesian coordinates which in 1D take the form
\beq
\phi_n(x)=\frac{1}{\sqrt{2^n n!}}\bigg(\frac{\omega}{\pi}\bigg)^{1/4}e^{-\omega x^2/2}H_n(\sqrt{\omega}x),
\eeq
where the $H_n$ are Hermite polynomials.
The sums over ${\bf k}_1$ and ${\bf k}_2$ in Eq.~(\ref{Eq:DQ11}) are independent and identical and take the form
\beq
\sum_{n=0}^\infty e^{-\beta E_{n} } |\phi_{n}(x)|^2 = \sqrt{\frac{\omega}{\pi}}e^{-\beta \omega /2-\omega x^2} 
G(\beta \omega,\sqrt{\omega},x),
\eeq
for each cartesian dimension, where the function $G$ can be calculated as a special case of
Mehler's formula~\cite{MehlerFormula}:
\begin{align*}
G(k,b,x) \equiv \sum_{n=0}^\infty \frac{e^{-k n}}{2^n n!}[H_n(b x)]^2 &= 
\frac{\exp [{2 b^2 x^2}/({1 + e^k})]}{\sqrt{1-e^{-2k}}}.
\end{align*}
This formula encodes the finite-temperature, single-particle density matrix of
a noninteracting, nonrelativistic system in a harmonic trapping potential, and therefore
its use is essential in the calculations that follow.

Squaring the result, we obtain, in $1$ spatial dimension,
\bea
\Delta Q_{11}&=&2C\int_0^\infty dx \frac{\omega}{2\pi\sinh(\beta \omega)}\exp\left[-2\omega x^2\tanh(\beta \omega/2)\right] \nonumber \\
&=& C\frac{\sqrt{\omega}}{4} \frac{1}{ \sqrt{\pi\sinh(\beta \omega)} \sinh(\beta \omega/2)},
\eea
where we have performed the last Gaussian integral along with some hyperbolic function simplifications.
Generalizing to $d$ spatial dimensions is very simple in this case:
\bea
\Delta Q_{11} &=& C \left[ \frac{\sqrt{\omega}}{4} \frac{1}{\sqrt{\pi\sinh(\beta \omega)} \sinh(\beta \omega/2)}\right]^d.
\eea
Using Eq.~(\ref{Eq:Q1}), we find that $Q_1$ cancels exactly in the final expression, as expected, such that
\beq
\label{Eq:Db2vsC}
\Delta b_2 = \frac{C}{2} \left[ \frac{\sqrt{\omega}}{2\sqrt{\pi\sinh(\beta \omega)}}\right]^d = 
\frac{C}{\lambda_T^d} \frac{1}{2}\left[ \frac{\beta \omega}{2 \sinh(\beta \omega)}\right]^\frac{d}{2},
\eeq
where we have used the thermal wavelength $\lambda_T = \sqrt{2\pi \beta}$ to write the result in
dimensionless form.

As mentioned above, we will use this result to connect to the physical coupling $\lambda$ of a given system, 
as a renormalization condition at a given value of $\beta \omega$. To that end, we first solve for $C/\lambda_T^d$:
\beq
\label{Eq:CvsDb2}
\frac{C}{\lambda_T^d} = 2\Delta b_2 \left[ \frac{2 \sinh(\beta \omega)}{\beta \omega}\right]^\frac{d}{2}.
\eeq
In the unitary limit of the 3D Fermi gas~\cite{ZwergerBook}, for instance, the exact answer for $\Delta b_2$ is known~\cite{VirialReview}:
\beq
\label{Eq:Db2BetaOmegaExact}
\Delta b_2 = \frac{1}{2} \left( \frac{e^{-\beta \omega / 2}}{1 + e^{- \beta \omega}} \right)= \frac{1}{4 \cosh(\beta \omega / 2)} .
\eeq
Using that result in Eq.~(\ref{Eq:CvsDb2}) yields
\beq
\frac{C}{\lambda_T^3} = \frac{e^{-\beta \omega/2}}{1 + e^{-\beta \omega}} 
\left[ \frac{{2 \sinh(\beta \omega)}}{{\beta \omega}}\right]^{3/2}.
\eeq
As we will show below, this type of renormalization prescription is very practical as our results for $\Delta b_3$ and 
$\Delta b_4$ are simple quadratic functions of ${C}/{\lambda_T^d}$ with $\beta\omega$-dependent coefficients.

\section{Results}

Following the steps outlined above in the example calculation of $\Delta b_2$, we
have calculated the various contributions to $\Delta b_3$ and $\Delta b_4$,
which we present in this section. In all cases, the central component of the calculation
is the use of the analytic form of Mehler's kernel, which effectively reduces the calculation to 
a small number of Gaussian integrals.

\subsection{Result for $\Delta Q_{2,1}$ and $\Delta b_3$}

With small modifications to the example for $\Delta b_2$, it is straightforward to show that 
\bea
\Delta Q_{2,1} &=& \frac{C}{\lambda_T^d} \left[\frac{{\beta\omega}}{{4\sinh^{3}(\beta \omega)}}\right]^\frac{d}{2} \times \nonumber \\
&& \!\!\!\!\!\!\!\!\!\!\!\!\! \left[ \frac{1}{2^\frac{d}{2}\tanh^d \left(\frac{\beta \omega}{2}\right)} - \left(\frac{2\cosh^2(\beta \omega/2)}{4\cosh^2(\beta \omega/2)-1}\right)^{{d}/{2}} \right],
\eea
and using the results of the previous section, it is easy to assemble the final answer for $\Delta b_3$
in our approximation using
\beq
\Delta b_3 = \frac{2 \Delta Q_{2,1}}{Q_1} - \Delta b_2 Q_1.
\eeq

Note that $Q_1$ diverges in the $\beta \omega \to 0$ limit, but that divergence will cancel out in the final expression
for $\Delta b_3$. Indeed, after simplifications, we obtain
\beq
\Delta b_3 = -\frac{C}{\lambda_T^d} \left[\frac{{\beta \omega}}{{2\sinh(\beta \omega)}}\right]^\frac{d}{2} \left(\frac{1}{4\cosh^2(\beta \omega/2)-1}\right)^{{d}/{2}},
\eeq
which is manifestly finite in the $\beta \omega \to 0$ limit. In that limit,
\beq
\Delta b_3 \to 
-\frac{C}{\lambda_T^d} \frac{1}{6^\frac{d}{2}}.
\eeq

We recall the result of Ref.~\cite{ShillDrut} for the homogeneous case, namely
\beq
\Delta b_3^{\text{hom}} = -\frac{C}{\lambda_T^d} \frac{1}{2^\frac{d}{2}}, 
\eeq
which shows that the relationship between the homogeneous and trapped cases, 
pointed out in the introduction, is also satisfied once interactions are turned on, as expected.

\subsection{Result for $\Delta Q_{3,1}$, $\Delta Q_{2,2}$, and $\Delta b_4$.}

Again following the steps outlined above, we obtain
\begin{widetext}
\bea
\frac{2 \Delta Q_{3,1}}{Q_1} &=& 
\frac{C}{2 \lambda_T^d}\left[\frac{\sqrt{2 \beta \omega}   \sinh(\beta \omega/2)   }{2 \sinh^2(\beta \omega)}\right]^d 
\times\left \{ 
\left(\frac{1}{2 \tanh ^{3}\left({\beta \omega}/{2}\right)}\right)^{\frac{d}{2}}  
+ 2 \tanh^\frac{d}{2}(\beta \omega)\left(\frac{\cosh(\beta \omega)+ 1}{2 \cosh(\beta \omega)+1} \right)^{\frac{d}{2}} 
\right . \nonumber \\
&&\ \ \ \ \ \ \ \ \ \ \ \left . -
2 \coth^\frac{d}{2}(\beta \omega /2)\left(\frac{\cosh(\beta \omega)+ 1}{2 \cosh(\beta \omega)+1} \right)^{\frac{d}{2}} 
- \left( \frac{1}{2\tanh \left({\beta \omega}/{2}\right)} \right)^{\frac{d}{2}} \right \} ,
\eea
\bea
\frac{\Delta Q_{2,2}}{Q_1} &=& 
\frac{C}{2 \lambda_T^d}\left[\frac{\sqrt{2 \beta \omega}  \sinh(\beta \omega/2)  }{2 \sinh^2(\beta \omega)}\right]^{d} 
\times \left\{
\left(\frac{1}{2 \tanh ^{3}\left({\beta \omega}/{2}\right)}\right)^{\frac{d}{2}} 
-2\coth^\frac{d}{2}({\beta \omega}/{2}) \left( \frac{\cosh(\beta \omega) + 1}{2 \cosh(\beta \omega) + 1}\right)^\frac{d}{2} 
+ \left(\frac{\tanh(\beta \omega)}{2}\right)^{\frac{d}{2}} 
\right . \nonumber \\
&&\ \ \ \ \ \ \ \ \ \ \ \left .
+\left(\frac{C}{2\lambda_T^{d}}\right)\frac{(2 \beta \omega)^\frac{d}{2}}{2^d}
\left[ 
\frac{1 }{\tanh^d\left({\beta \omega}/{2}\right)}+1-2\left(\sech(\beta \omega)+1\right)^\frac{d}{2}
\right]\right\}.
\eea
\end{widetext}

Combining these with results from the previous sections, the final answer for $\Delta b_4$ can be assembled using  
\bea
\!\Delta b_4\!=\! \frac{2 \Delta Q_{3,1} \!+\! \Delta Q_{2,2}}{Q_1}\!-\! \Delta\!\left(b_3 + \frac{b_2^2}{2}\right)\! Q_1\!-\!\frac{\Delta b_2}{2} Q_1^2.
\eea

After several simplifications and cancellations (which can be tracked by their degree of divergence as $\beta \omega \to 0$, while the
final result for $\Delta b_4$ is finite), we obtain
\bea
\Delta b_4 &=& \frac{C}{2\lambda_T^d}\left[\frac{\sqrt{2 \beta \omega}  \sinh(\beta \omega/2)  }{2 \sinh^2(\beta \omega)}\right]^{d}
\times \nonumber \\
&&\!\!\!\!\!\!\!\!\! \left \{ 
2\tanh^\frac{d}{2}(\beta \omega)\left(\frac{\cosh(\beta \omega)+ 1}{2 \cosh(\beta \omega)+1} \right)^{\frac{d}{2}} + \left(\frac{\tanh(\beta \omega)}{2}\right)^{\frac{d}{2}} \nonumber \right .\\
&&\!\!\!\!\!\!\!\!\!+
\left .
\left(\frac{C}{2\lambda_T^{d}}\right)
\frac{(2 \beta \omega)^\frac{d}{2}}{2^d}
\left[
1 - 2 \left(\sech(\beta \omega)+1\right)^\frac{d}{2}
\right]
\right \}.
\eea

In this case, the limit $\beta \omega \to 0$ yields 
\bea
\Delta b_4 &\to& \frac{C}{\lambda_T^d}  4^{-\frac{d}{2}}(3^{-\frac{d}{2}} + 2^{-d-1} )\nonumber \\
&& + 
\left( \frac{C}{\lambda_T^d} \right)^2 4^{-d-1} \left( 1 - 2^{\frac{d}{2} +1} \right).
\eea

Once again, we recall the homogeneous result:
\bea
\Delta b_4^\text{hom} &=& \frac{C}{\lambda_T^d}(3^{-\frac{d}{2}}\!\! +\! 2^{-d-1} ) \nonumber \\
&&+ \left( \frac{C}{\lambda_T^d} \right)^2 4^{-\frac{d}{2}-1}(1 - 2^{\frac{d}{2}+1}) ,
\eea
and find that $\Delta b_4 = \Delta b_4^\text{hom} 4^{-\frac{d}{2}}$, as expected.

\subsection{Results in terms of $\Delta b_2$.}

Finally, collecting our results for $\Delta b_3$ and $\Delta b_4$ and expressing them in terms of $\Delta b_2$
[via Eq.~(\ref{Eq:CvsDb2})], we obtain
\beq
\Delta b_3 = - 2\left(\frac{1}{4\cosh^2(\beta \omega/2)-1}\right)^\frac{d}{2} \Delta b_2,
\eeq
and
\beq
\Delta b_4 =
\left(\frac{1}{2\cosh(\beta \omega)}\right)^\frac{d}{2} \!\!\!\! f_1(\beta \omega, d) \Delta b_2 + f_2(\beta \omega, d) (\Delta b_2)^2,
\eeq
where
\beq
f_1(\beta \omega, d) = \left(\frac{ 1 }{ 2\cosh(\beta \omega/2)}\right)^{d} + 2\left(\frac{1}{ 4\cosh^{2}(\beta \omega/2)-1} \right)^\frac{d}{2},
\eeq
and
\beq
f_2(\beta \omega, d) = \left(\frac{ 1 }{ 2\cosh(\beta \omega/2)}\right)^{d} - 2 \left(\frac{ 1 }{ 4\cosh^2(\beta \omega/2) - 2}\right)^\frac{d}{2}.
\eeq

The above formulas for $\Delta b_3$ and $\Delta b_4$ are the main result of this work. In the following we explore their behavior as a function of 
$d$ and $\beta \omega$, focusing in particular on the unitary limit of the 3D Fermi gas. While numerical results 
exist for these quantities in some cases, in particular in 2D~\cite{DrummondVirial2D, virial2D2, PhysRevA.89.013614, Ngampruetikorn} (see also~\cite{virial2D, Daza2D, Ordo}) and in 3D at unitarity~\cite{LeeSchaeferPRC1, LiuHuDrummond, LiuHuDrummond2, DBK}, most of those correspond to homogeneous systems and do not 
feature explicit, analytic dependence on the dimension nor on $\beta \omega$, as shown here. Our results are therefore useful in that they are 
able to provide analytic insight into the behavior of virial coefficients across dimensions, and as a function of the temperature (or trapping frequency)
as well as the coupling strength. Below, we evaluate our formulas and discuss the resulting answers.

\subsection{Qualitative behavior.}

\begin{figure}[t]
  \begin{center}
  \includegraphics[scale=0.54]{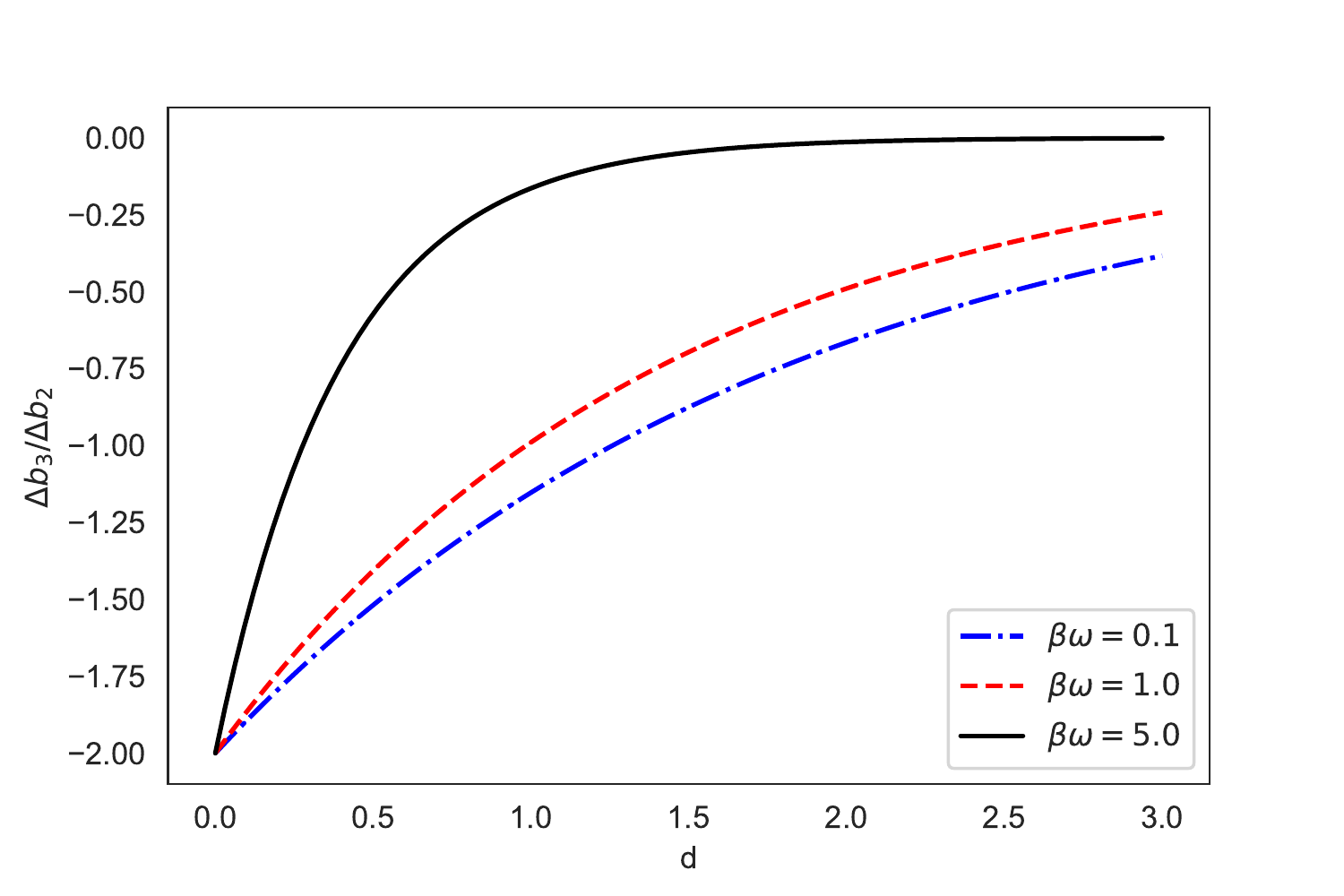}
  \includegraphics[scale=0.54]{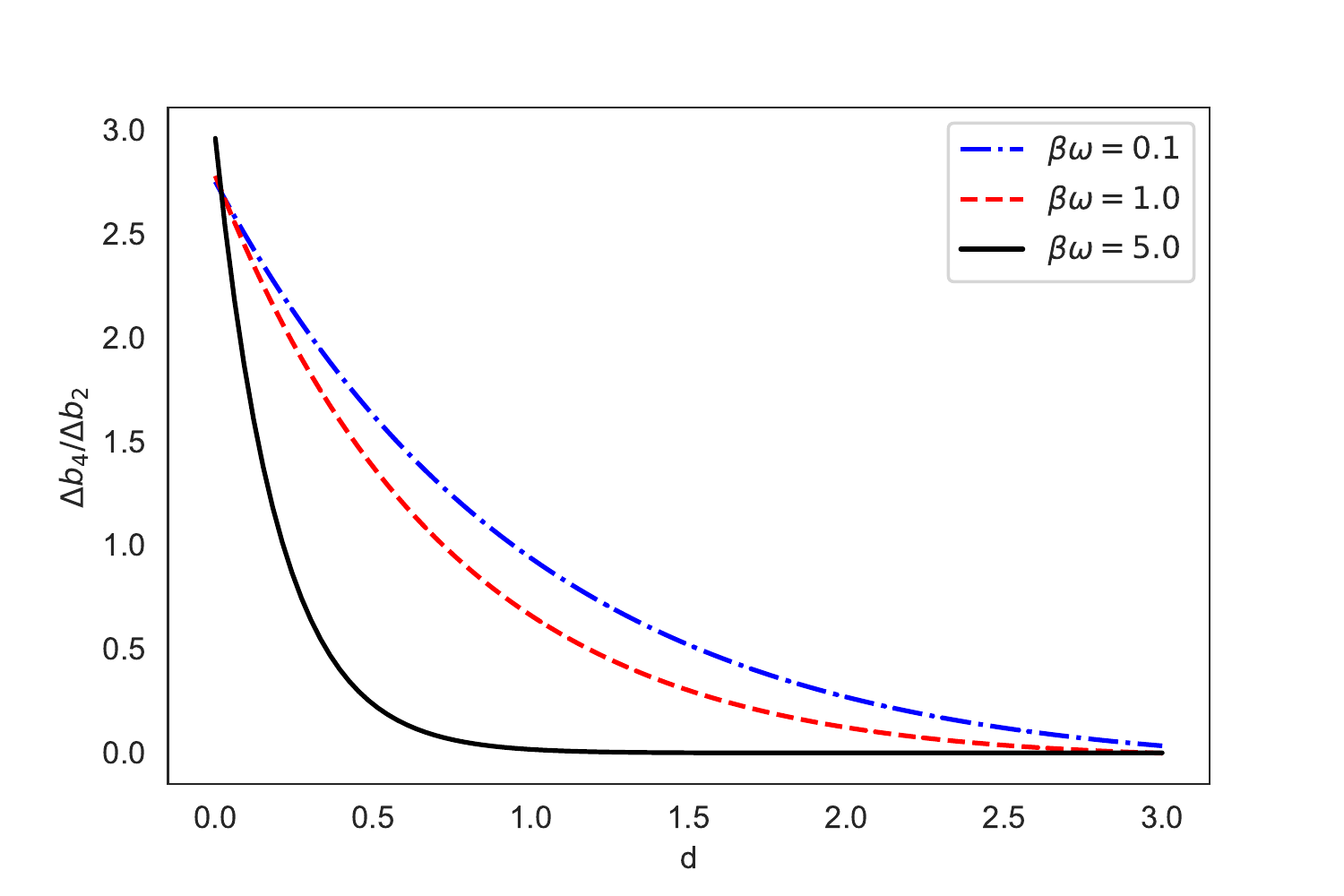}
  \end{center}
  \caption{$\Delta b_3/\Delta b_2$ (top) and $\Delta b_4/\Delta b_2$ (bottom) as functions of the spatial dimension $d$
  at fixed $\beta \omega = 0.1$, $1.0$, and $5.0$, for $\Delta b_2$ corresponding to the unitary limit in $d=3$, Eq.~(\ref{Eq:Db2BetaOmegaExact}).
   }
  \label{Fig:Db3etDb4vsDb2}
\end{figure}

To illustrate our analytic results, in Fig.~\ref{Fig:Db3etDb4vsDb2} we show $\Delta b_3/\Delta b_2$ and $\Delta b_4/\Delta b_2$ as a function of 
the spatial dimension $d$, at various $\beta \omega$, fixing $\Delta b_2$ to its value in the unitary limit (as a reference point).
We find that, as $d$ increases, the magnitude of the interaction-induced change $\Delta b_n$ decreases. This suggests that, using $\Delta b_2$ 
as the fixed, dimension-independent coupling, the radius of convergence of the virial expansion increases with $d$. This is consistent with the
idea that, in higher dimensions, the kinetic energy dominates over the interactions and mean-field type of approaches capture the behavior of
the system correctly.

\begin{figure}[t]
  \begin{center}
  \includegraphics[scale=0.54]{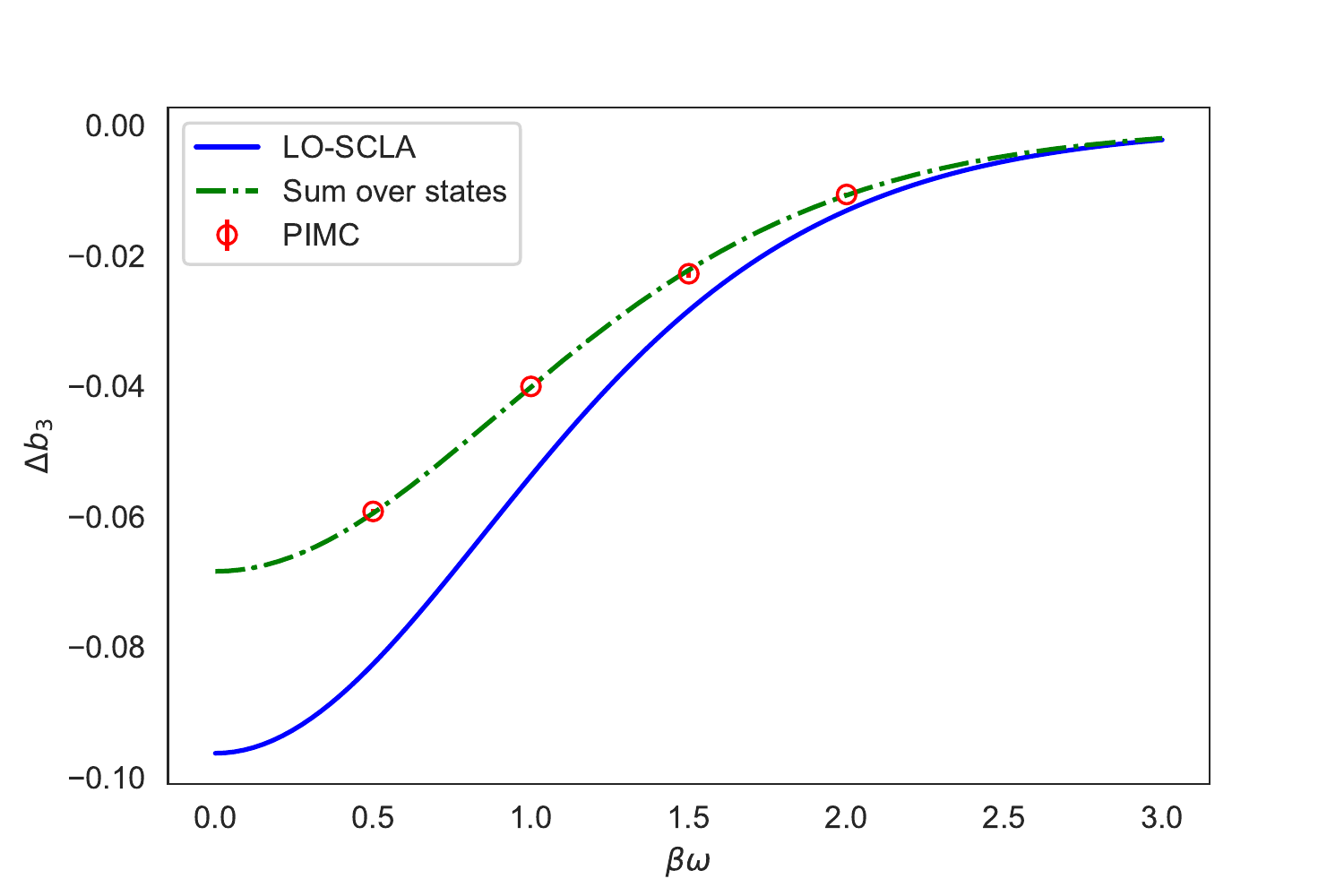}
  \includegraphics[scale=0.54]{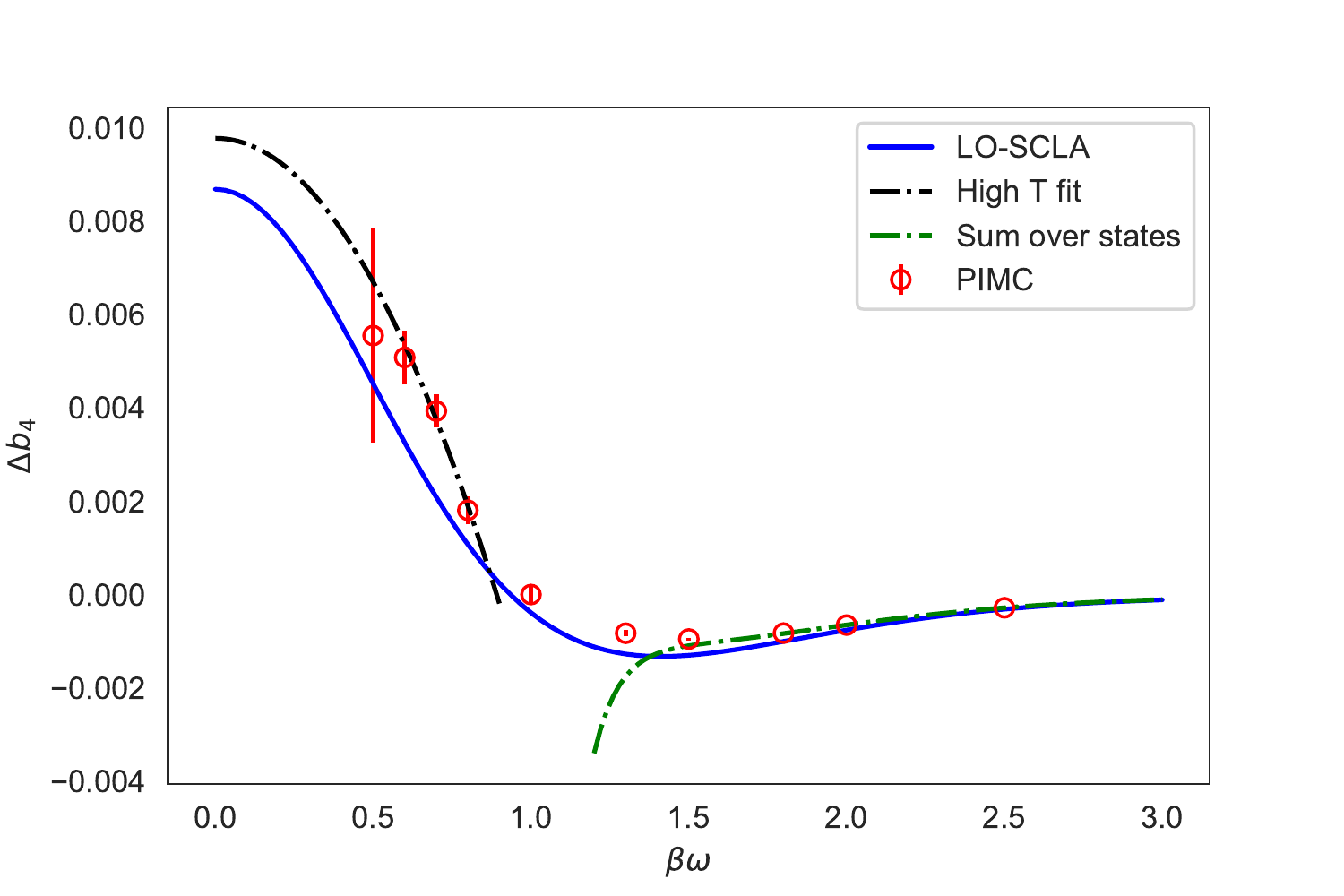}
  \end{center}
  \caption{Comparison of our LO-SCLA results for $\Delta b_3$ (top) and $\Delta b_4$ (bottom) as a function of $\beta \omega$.
  The high-temperature fits and PIMC results are from Ref.~\cite{Doerte}; the sum-over-states results are from Ref.~\cite{Rakshit}.
   }
  \label{Fig:ComparisonYanBlume}
\end{figure}

As a comparison with previous calculations, we show in Fig.~\ref{Fig:ComparisonYanBlume} our results
in 3D at unitarity as a function of $\beta \omega$, superimposed with the data from Ref.~\cite{Doerte}.
While we do not expect, a priori, good quantitative agreement in this strong coupling regime, we
find at least qualitative agreement for both $\Delta b_3$ and $\Delta b_4$, and
surprisingly good agreement at the quantitative level for $\Delta b_4$. Clearly, the LO-SCLA is able
to capture more than just the shape of the $\beta \omega$ dependence of the virial coefficients.

\section{Summary and Conclusions}

In this work we have implemented a semiclassical approximation, at leading order, to calculate the virial coefficients
$\Delta b_3$ and $\Delta b_4$ of harmonically trapped Fermi gases. Our calculations yield analytic answers 
as functions of $\beta \omega$ and, by a renormalization prescription that matches $\Delta b_2$ to the known exact result, 
we also obtain the dependence on the physical coupling strength. Notably, our results are also analytic functions of the spatial dimension $d$,
allowing us to study the behavior of the virial expansion across dimensions. We find that, at fixed $\Delta b_2$, the magnitude of 
$\Delta b_n$ decreases as $d$ increases, for all $\beta \omega$. 

Although there have been many (and very precise) determinations of virial coefficients in the literature, they are mostly numerical
and focus on specific dimensions or couplings (and most of them are for homogeneous systems).
Our approach and results are, in that sense, complementary: we do not expect high precision from the LO-SCLA, but through it we are 
able to study, explicitly, the variations with the parameters of the problem, which yield qualitative analytic insight into the properties
of the virial expansion. We have demonstrated the quality of our leading-order (!) results for $\Delta b_3$ and $\Delta b_4$ in the unitary limit 
by showing that they qualitatively follow the expected answers, which is an encouraging sign to proceed to next-to-leading order in future work.
Furthermore, the approximate agreement with prior results at unitarity suggests that, between the noninteracting regime and the unitary point, 
that agreement should be even better than shown here.

Finally, it should be pointed out that the renormalization prescription based on $\Delta b_2$ does not by itself eliminate all the lattice 
artifacts. Future studies should explore the use of improved actions (see e.g.~\cite{DrutNicholson, Drut}), potentially making use of
prior knowledge of $\Delta b_3$ were available, to enhance the quality of the expansion.


\acknowledgments
This material is based upon work supported by the National Science Foundation under Grant No.
PHY{1452635} (Computational Physics Program).
C.E.B. acknowledges support from the United States Department of Energy through the
Computational Science Graduate Fellowship (DOE CSGF) under grant number DE-FG02-97ER25308.


\end{document}